\documentclass{article}
\usepackage{latexsym}
\usepackage{graphics}
\usepackage{color}
\newtheorem{proposition}{Proposition}
\newtheorem{theorem}{Theorem}
\newtheorem{lemma}{Lemma}

\begin{document}
\title{Spherically symmetric spacetimes with a trapped surface}
\author{Mihalis Dafermos}
\maketitle
\begin{abstract}
This paper investigates the
global properties of a class of spherically symmetric spacetimes.
The class contains the maximal 
development of asymptotically flat spherically symmetric
initial data for a wide variety of coupled Einstein-matter systems. 
For this class, it is proven here 
that the existence of a single trapped or marginally 
trapped surface implies the completeness of future null infinity 
and the formation of an event horizon whose area radius is bounded 
by twice the final Bondi mass. 
\end{abstract}
One of the fundamental questions in gravitational collapse is
the so-called \emph{weak cosmic censorship} conjecture~\cite{chr:givp,rp:sst}. 
This is the statement
that, for generic asymptotically flat initial data, solutions to appropriate
Einstein-matter systems possess a 
complete null infinity. For a precise definition of
this latter concept, the reader should consult~\cite{chr:givp}. 

In~\cite{chr:ins}, Christodoulou proves weak cosmic
censorship for the collapse of a spherically symmetric self-gravitating
scalar field. His argument proceeds by showing that data leading to a
naked singularity satisfy the property--when perturbed generically--that
all singularities are ``preceeded'' by trapped surfaces.
The completeness of null infinity is then inferred from this property.

In the present paper, we formulate general assumptions which,
in the context of spherical symmetry, ensure that the existence
of a \emph{single} trapped surface suffices to show the formation of a black 
hole and the completeness of null infinity. 
The most restrictive assumption excludes a certain kind of TIP
not emanating from the center. The assumption has been 
shown to hold for the maximal development
of asymptotically flat data for a wide variety
of Einstein-matter systems. Indeed, in this context the assumption
corresponds to the statement that ``first singularities'' arising from non-trapped
points can only emanate from the center. 
For the systems for 
which the assumptions here hold, the results of this paper 
suggest a local approach to proving weak cosmic censorship.

Finally, we note that in 
the process of proving the completeness of null infinity,
we obtain an upper bound of twice the final Bondi 
mass for the area radius of the apparent horizon and--more interestingly--the
\emph{event horizon} 
of the black hole that forms.  Upper bounds of this
form are commonly known as ``Penrose inequalities''.

\section{First assumptions}
Our assumptions are motivated from properties of the maximal developments
of spherically symmetric asymptotically flat
initial data, for ``appropriate'' Einstein-matter systems.
We will formulate assumptions directly at the level
of a two dimensional submanifold $\mathcal{Q}^+$ of $2$-dimensional
Minkowski space, endowed with metric $-\Omega^2dudv$, 
a function $r$, and a symmetric two-tensor $T_{ab}$. In applications,
$\mathcal{Q}^+$ arises as the quotient of a spherically symmetric
maximal development by the group $SO(3)$. In general, we shall include
a discussion of the realm
of applicability of each assumption--and any subtleties that
might arise--immediately after its formulation.

\subsection{The quotient manifold}
Let ${\bf R}^2$ denote the standard plane.
We will call its coordinates $(u,v)$, and we will depict
the $v$-axis at $45$ degrees from the horizontal, and the $u$-axis
at $135$ degrees. Unless otherwise noted, causal-geometric concepts,
like the word ``timelike'' or the set $J^+(p)$, etc., will refer to the 
metric $-dudv$ of ${\bf R}^2$, future oriented in the standard way
so that $u$ and $v$ are both increasing towards the future. 
Our first assumption is
\begin{enumerate}
\item[${\bf A}'$]
We are given a bounded 
two-dimensional submanifold $\mathcal{Q}^+\subset{\bf R}^2$ with
boundary $\Gamma\cup S$, where $\Gamma$ is a connected timelike curve, 
and $S$ is a connected spacelike curve, 
and $\Gamma \cap S$ is a single point $p$.
We assume that on $\mathcal{Q}^+$ we are given 
$C^1$ functions $r$, $\Omega$, such that $\Omega>0$, $r\ge0$,
and $r(q)=0$ iff $q\in \Gamma$. 
Defining the so-called $\emph{Hawking mass}$:
\[
m=\frac r2(1+4\Omega^{-2}\partial_ur\partial_vr),
\]
we assume that $m$ is uniformly bounded along $S$.

We assume that $\mathcal{Q}^+$ is foliated by connected constant
$v$-segments with past endpoint on $S	
$, and also by
connected constant $u$-segments with past endpoint on $\Gamma\cup S$. 
We will call the former ``ingoing'' segments, and the ``latter'' outgoing.
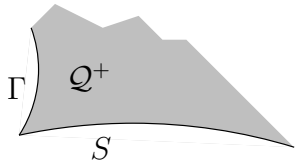
\[
\input{A.pstex_t}
\]
\end{enumerate}

Given an ``appropriate'' notion of spherically
symmetric asymptotically flat dataset $(\Sigma, \bar{g}, K, \ldots)$ for
a ``reasonable'' Einstein-matter system, with one end, it follows that
the maximal Cauchy development\footnote{``reasonable'' means in particular
that this concept can be defined with its standard properties...}
$(\mathcal{M},g)$ will admit an isometry action by the group
$SO(3)$ and moreover, $\mathcal{Q}=\mathcal{M}/SO(3)$ will inherit\footnote{
``appropriate'' is taken to ensure this\ldots}
the structure of a Lorentzian manifold.  If $\mathcal{Q}^+$ denotes
the quotient of $\mathcal{M}\cap J_{g}^+(\Sigma)$, then $\mathcal{Q}^+$ can be
conformally embedded into ${\bf R}^2$ so as to satisfy
the conditions of ${\bf A}'$, where $-\Omega^2dudv$
represents the metric of $\mathcal{Q}$, and $-\Omega^2dudv+r^2\gamma$
represents the metric of $\mathcal{M}$, where $\gamma$ is the standard
metric on $S^2$. $\Gamma$, the center, corresponds to the
set of fixed points of the group action.

\subsection{Structure equations}
Our next assumption is
\begin{enumerate}
\item[${\bf B}'$]
We have a symmetric $2$-tensor,
with bounded components $T_{uu}$, $T_{uv}$, $T_{vv}$,
defined on $\mathcal{Q}^+$.
The following equations hold pointwise almost everywhere:
\begin{equation}
\label{gia-to-u}
\partial_u(\Omega^{-2}\partial_ur)=-4\pi r\Omega^{-2}T_{uu}
\end{equation}
\begin{equation}
\label{gia-to-v}
\partial_v(\Omega^{-2}\partial_vr)=-4\pi r\Omega^{-2}T_{vv},
\end{equation}
\begin{equation}
\label{m-u}
\partial_um={8\pi r^2}\Omega^{-2}(T_{uv}\partial_ur-
T_{uu}\partial_vr),
\end{equation}
\begin{equation}
\label{m-v}
\partial_vm={8\pi r^2}\Omega^{-2}(T_{uv}\partial_vr-
T_{vv}\partial_ur).
\end{equation}
\end{enumerate}

If the metric $g$  defined by $-\Omega^2dudv+r^2\gamma$ 
is $C^2$, then equations $(\ref{gia-to-u})$--$(\ref{m-v})$
are just identities necessarily satsified by the $uu$, $uv$ and $vv$
components of the tensor
\[
T_{\mu\nu}=\frac{1}{8\pi}(R_{\mu\nu}-\frac12g_{\mu\nu}R).
\]
In the context of spherical symmetry, one often defines
less regular notions of solutions tied explicitly to 
$(\ref{gia-to-u})$--$(\ref{m-v})$. It is for this reason that
we have prefered to postulate $(\ref{gia-to-u})$--$(\ref{m-v})$ directly.

\subsection{Positive energy condition}
\begin{enumerate}
\item[${\bf \Gamma}'$]
We have
\begin{equation}
\label{energyc}
T_{uu}\ge0, T_{vv}\ge0, T_{uv}\ge0
\end{equation}
\end{enumerate}
Again, when $T_{\mu\nu}$ is the energy momentum tensor
of matter, coupled to gravity via the Einstein equations, 
the above assupmtion is precisely
the positive energy condition.

\subsection{No anti-trapped surfaces initially}
\begin{enumerate}
\item[${\bf \Delta}'$]
We have
\[
\partial_ur < 0
\]
along $S$.
\end{enumerate}

This condition has been introduced by Christodoulou in~\cite{chr:sgrf}.
It is motivated, in part, by Proposition~\ref{toprwto} of the next section.

\section{$\mathcal{R}$,$\mathcal{T}$, and $\mathcal{A}$}
In what follows we assume ${\bf A}'$--${\bf \Delta}'$.

Following the notation of~\cite{chr:sgrf}, we define
the \emph{regular} or \emph{non-trapped region} 
\[
\mathcal{R}=\{p\in{\mathcal{Q}^+}{\rm\ such\ that\ }
\partial_vr>0, \partial_ur<0\},
\]
the \emph{trapped region} 
\[
\mathcal{T}=\{p\in{\mathcal{Q}^+}{\rm\ such\ that\ }
\partial_vr<0, \partial_ur<0\},
\]
and the \emph{marginally trapped region}
\[
\mathcal{A}=\{p\in{\mathcal{Q}^+}{\rm\ such\ that\ }
\partial_vr=0, \partial_ur<0\}.
\]
We include for completeness the proof of the following
proposition, due to Christodoulou~\cite{chr:sgrf}:
\begin{proposition}
\label{toprwto}
We have $\mathcal{Q}^+=\mathcal{R}\cup\mathcal{T}\cup\mathcal{A}$,
i.e.~anti-trapped surfaces are non-evolutionary.
On $\mathcal{A}$, $1-\frac{2m}r=0$, while $1-\frac{2m}r<0$ in $\mathcal{T}$. In
$\mathcal{R}$, we have $m\ge0$ and $\partial_vm\ge0$, $\partial_um\le0$.
Moreover, if $(u,v)\in{\mathcal{T}}$, then $(u,v^*)\in\mathcal{T}$
for $v^*>v$, and similarly, if $(u,v)\in\mathcal{T}\cup\mathcal{A}$,
then $(u,v^*)\in\mathcal{T}\cup\mathcal{A}$. 
\end{proposition}
\noindent\emph{Proof.} 
By assumption ${\bf A}'$, 
all ingoing curves in $\mathcal{Q}$ have past endpoint on $S$. 
Thus, integrating $(\ref{gia-to-u})$
from $S$, we obtain--in view of assumptions ${\bf \Gamma}'$
and ${\bf \Delta}'$--that $\Omega^{-2}\partial_ur<0$, and thus
$\partial_ur<0$ in $\mathcal{Q}^+$. This proves the first statement.

The second statement is an immediate consequence of the identity
\[
1-\frac{2m}r=-\frac{4}{\Omega^2}\partial_ur\partial_vr,
\]
in view of the inequality $\partial_ur<0$.

Integrating now $(\ref{gia-to-v})$ yields that $\Omega^{-2}\partial_vr$
is a nonincreasing function of $v$, and this immediately yields the
final statement.

For the third statement, note first that the 
inequalities $\partial_u m\le0$, $\partial_v m\ge 0$, on $\mathcal{R}$,
are trivial consequences of the signs of $\partial_vr$ and $\partial_ur$
in $(\ref{m-u})$ and $(\ref{m-v})$,
in view of Assumption ${\bf \Gamma}'$.
To show that $m\ge 0$ on $\mathcal{R}$, in view of the fact that the statement
proved in the previous paragraph shows that
$(u,v)\in\mathcal{R}$ implies $(u,v^*)\in\mathcal{R}$ for all $v^*\le v$,
it suffices to show that $m\ge0$ on $(\Gamma\cup S)\cap\mathcal{R}$.

The condition $m=0$ on $\Gamma$ is implied by the regularity assumption
of ${\bf A}'$. Let ${\bf K}$
denote the unit tangent vector on $S$ such that ${\bf K}\cdot v>0$.
It follows from $(\ref{m-v})$, $(\ref{m-u})$, that ${\bf K}\cdot m\ge0$
on $S\cap\mathcal{R}$. Let $s$ denote the coordinate on $S$ with 
${\bf K}\cdot s=1$, $s=0$ at $\Gamma\cap S$. 
If $s'\in S\cap\mathcal{R}$ then either
$[0,s')\in S\cap\mathcal{R}$, in which case $m(s')\ge m(0)=0$,
or else $(t',s')\subset\mathcal{R}$ for $t'\in\mathcal{A}$,
in which case $m(s')\ge m(t')$. But $m(t')=\frac{r(t')}2>0$.
This completes the proof.
$\Box$

\section{Null infinity}
The curve $S$ acquires a unique limit point $i_0$ in 
$\overline{\mathcal{Q}^+}\setminus\mathcal{Q}^+$ 
called \emph{spacelike infinity}. 
Let $\mathcal{U}$ denote the set of all $u$
defined by
\[
\mathcal{U}=\left\{u: \sup_{v:(u,v)\in\mathcal{Q}^+} r(u,v)=\infty\right\}.
\]
This set may of course be empty, even if $r\to\infty$ along $S$.
For each $u\in\mathcal{U}$, there is clearly a unique $v^*(u)$
such that 
\[
(u,v^*(u))\in\overline{\mathcal{Q}^+}\setminus\mathcal{Q}^+.
\]
Define
\[
\mathcal{I}^+=\bigcup_{u\in\mathcal{U}}(u,v^*(u))
\] 
We will call $\mathcal{I}^+$
 \emph{future null infinity}. 
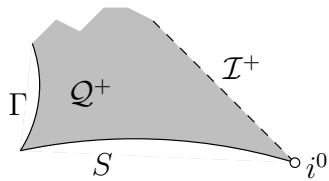
\[
\input{nullinfinity.pstex_t}
\]
We have
the following
\begin{proposition}
If non-empty, $\mathcal{I}^+$ is a connected
ingoing null ray with past limit point $i_0$.
\end{proposition}
\noindent\emph{Proof}.
Let $i_0=(U,V)$. If $v_0<V$, then since $\partial_ur<0$ in $\mathcal{Q}^+$, 
we have
an a priori bound for $r$ in $\mathcal{Q}^+\cap\{v\le v_0\}$
by the supremum of $r$ on $\{v\le v_0\}\cap S$. 
Thus $\mathcal{I}^+\cap\{v=v_0\}=\emptyset$, i.e., 
$\mathcal{I}^+\subset\{v=V\}$. 

Suppose now that $(u_0,V)\in\mathcal{I}^+$ and let $u<u_0$.
Since, by definition $\lim_{v\to V} r(u_0,v)=\infty$,
while on the other hand $r(u,v)> r(u_0,v)$ by the 
inequality $\partial_u r<0$, it follows that $\lim_{v\to\infty} r(u,v)=\infty$,
i.e., $(u,V)\in\mathcal{I}^+$. This proves the proposition.
$\Box$

We introduce the assumption:
\begin{enumerate}
\item[${\bf E}'$]
$\mathcal{I}^+$ is non-empty.
\end{enumerate}
In applications to the initial value problem, 
for initial data such that the matter is of compact support (or electrovacuum outside
a compact set)--and
such that the cosmological constant vanishes!--this assumption is
immediate by Birkhoff's theorem and the domain of dependence property.
It can also be reasonably expected to hold for matter whose initial
asymptotic behavior is sufficiently tame.

The set
$J^-(\mathcal{I}^+)\cap \mathcal{Q}^+$ 
is the so-called \emph{domain of outer communications.}
Clearly, by Proposition~\ref{toprwto}, it follows that 
\begin{equation}
\label{tae3w}
J^-(\mathcal{I}^+)\cap\mathcal{Q}^+\subset \mathcal{R}.
\end{equation}
From the inequalities $\partial_vr\ge0$, $\partial_ur\le0$ in $\mathcal{R}$,
and its uniform boundedness intially by ${\bf A}'$,
it is clear that $m$ extends to a nonincreasing
non-negative function along $\mathcal{I}^+$. We 
will denote $\inf_{\mathcal{I}^+}m$
by $M_f$, and refer to this as the \emph{final Bondi mass}.

\section{The extension principle}
To proceed further we will need one final assumption:

\begin{enumerate}
\item[${\bf \Sigma T}'$]
Let $p\in\overline\mathcal{R}\setminus\overline\Gamma$, 
and $q\in\overline{\mathcal R}\cap I^-(p)$ 
such 
that such that $J^-(p)\cap J^+(q)\setminus\{p\}\subset\mathcal{R}\cup\mathcal{A}$:
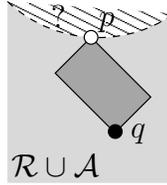
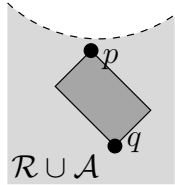
\[
\input{extension.pstex_t}
\]
Then $p\in\mathcal{R}\cup\mathcal{A}$.
\[
\input{extension2.pstex_t}
\]
\end{enumerate}
In the evolutionary context,
this assumption can be stated informally as the proposition that
a ``first singularity'' emanating from the regular region can only
arise from the center. It has been proven to hold
for a wide class of self-gravitating Higgs' fields~\cite{md:ns}
and for self-gravitating collisionless matter~\cite{rrs:reg,ar:in,dr:ip}. 
It can reasonably be expected to hold for charged scalar fields,
self-gravitating sigma models,
Yang-Mills fields, and more complicated systems arising from the coupling
of all the aforementioned. The reader should note, however, that, as applied to
maximal developments, ${\bf \Sigma T}'$ is more
restrictive than the previous assumptions, as it captures
a non-trivial feature of the behavior of solutions to the p.d.e.,
and not just ``general theory''. For instance,
${\bf \Sigma T}'$ is violated for a self-gravitating dust.


There are various equivalent ways of formulating ${\bf \Sigma T}'$.
Let
 $\mathcal{Q}^*$ denote the intersection of $\overline{\mathcal{Q}^+}$
with the set $\{v\ne V\}$.
Since $r$ is decreasing on ingoing null rays, 
$r$ can be extended by monotonicity to a function defined on 
$\mathcal{Q}^*$.
In view of the fact that a $p$ satisfying ${\bf \Sigma T}'$
is necessarily in $\mathcal{Q}^*$, one can replace
the assumption
$p\not\in\overline\Gamma$ with the assumption $p\in\mathcal{Q}^*$ and
$r(p)>0$.
 
Assuming ${\bf \Sigma T}'$, 
we can proceed to describe $\overline\mathcal{R}\setminus\mathcal{Q}^+$.
One can easily deduce from ${\bf \Sigma T}'$ and Proposition~\ref{toprwto} that if
$p\in \overline\mathcal{R}\setminus\mathcal{Q}^+$, then either
$p$ is on the outgoing null ray emanating from the future limit point
of $\Gamma$, or on the ingoing null ray emanating from the future limit
point, call it $i^+$, of $\mathcal{I}^+$.\footnote{Note 
that \emph{a priori}, as defined, $i^+$ may or may not be
contained in $\mathcal{I}^+$.} Let 
us denote by $\mathcal{B}_0$ the former ray, intersected 
with $\mathcal{Q}^*$, and by $\mathcal{C}^+$, the latter
ray intersected with $\overline\mathcal{Q}^+$, where this
latter ray is taken \emph{not} to include $i^+$. 
The set $\overline\mathcal{R}\setminus\mathcal{Q}^+$ is then 
the union of $\mathcal{I}^+$, $i^+$, and connected
closed subsets of
$\mathcal{B}_0$ and $\mathcal{C}^+$.
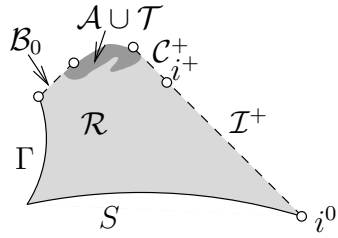
\[
\input{global.pstex_t}
\]
Of course, $\mathcal{B}_0$ may be a single point,
and $\mathcal{C}^+$ may indeed be empty.
Moreover, $i^+$ and $\mathcal{B}_0$ may coincide, cf.~Minkowski
space.


\section{The completeness of null infinity}

In what follows, we assume ${\bf A}'$--${\bf \Sigma T}'$.
The main result of this paper is
\begin{theorem}
If $\mathcal{A}$ is non-empty, then 
$\mathcal{I}^+$ is future complete.
\end{theorem}
By the statement that ``$\mathcal{I}^+$ is future complete'', we mean
that the spacetime $J^-(\mathcal{I}^+)\cap\mathcal{Q}^+\times S^2$ 
with respect to the metric
$-\Omega^2dudv +r^2\gamma$ satisfies the 
formulation of Christodoulou described in~\cite{chr:givp}.
\vskip1pc
\noindent\emph{Proof.} 
Since $\mathcal{A}\cup\mathcal{T}$ is non-empty, it follows 
by $(\ref{tae3w})$ that $J^-(\mathcal{I}^+)\cap \mathcal{Q}^+$
has a future boundary in $\mathcal{Q}^+$. 
This future boundary is an outgoing null ray
$\mathcal{H}$ we shall call the \emph{event horizon}. 
It is clear that
\begin{equation}
\label{profavws}
\mathcal{H}\subset\mathcal{R}\cup\mathcal{A}.
\end{equation}
Recall $\mathcal{Q}^*$ defined earlier.
By ${\bf \Sigma T}'$, we have
$\mathcal{H}=\mathcal{Q}^*\cap\{u=\tilde{U}\}$
for some $\tilde{U}$, i.e., $\mathcal{H}$ cannot terminate before reaching
 $i^+$. 
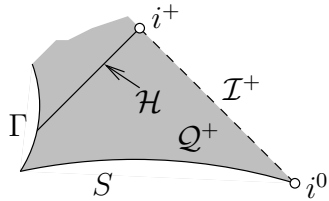
\[
\input{geg.pstex_t}
\]

\subsection{Outermost apparent horizon}
Define the set
\[
\mathcal{A}'=\{(u,v)\in\mathcal{A}:(u^*,v)\in\mathcal{R}
{\rm\ for\ all\ } u^*<u\}.
\]
We will call this set the \emph{outermost apparent horizon}.
First we show 
\begin{lemma}
\label{faivomevo}
$\mathcal{A}'$ is a non-empty (not necessarily connected)
achronal curve intersecting
all ingoing null curves for $v\ge v_0$, for sufficiently large $v_0$.
\[
\input{faiv.pstex_t}
\]
\end{lemma}
\noindent\emph{Proof.}
The statement that $\mathcal{A}'$ is achronal is immediate in view
of its definition and Proposition~\ref{toprwto}.
Let $(u',v')\in\mathcal{A}\cup\mathcal{T}$.
Let $v_0>v'$ be such that $S\cap\{v\ge v_0\}\subset\mathcal{R}$, and
consider an ingoing null ray $v=v''$ emanating from a point 
$(u'',v'')\in S$, where $v''\ge v_0$. If
$[u'',u']\times v''\in\mathcal{Q}$, then by Proposition~\ref{toprwto}, it
follows that $(u',v'')\in\mathcal{T}\cup\mathcal{A}$,
and thus, there must exist a point $(u^*,v'')\in\mathcal{A}'$.
In general, if $(\tilde{u},v'')\in\mathcal{T}\cup\mathcal{A}$, then
again there must exist a point $(u^*,v'')\in\mathcal{A}'$.
Thus, we must exclude the possibility that 
$[u',\bar{u})\times v''\in\mathcal{R}$ for some $\bar{u}$, with 
$(\bar{u},v'')\in\overline\mathcal{R}\cap\mathcal{Q}^*\setminus\mathcal{Q}$.
But this would imply that $(u^*,v^*)\in\mathcal{R}$ for
all  $u^*<\bar{u}$,
$v^*\le v''$. Since $\bar{u}\le u'$, it follows that the outgoing
null curve $u=\bar{u}$ intersects $\mathcal{R}\cup\mathcal{A}$. 
In particular, there is a point on this curve such that $r\ge c>0$.
Thus, it follows from $\partial_v r\ge0$ in $\mathcal{R}\cup\mathcal{A}$
that $r(\bar{u},v'')\ge c$, and thus, by ${\bf \Sigma T}'$,
$(\bar{u},v'')\in\mathcal{Q}$, a contradiction. $\Box$

Note that if $\mathcal{C}^+\cap\mathcal{B}_0\neq\emptyset$, then
it is \emph{not} necessarily the case that arbitrarily late
outgoing null curves enter $\mathcal{A}\cup\mathcal{T}$.

\subsection{Penrose inequality for the apparent horizon}
We have in fact the following
\begin{lemma} 
\label{faivomevopenrose}
On $\mathcal{A}'$,
$r\le 2M_f$.
\end{lemma}
\noindent The above lemma is one manifestation of what is commonly
referred to in the literature as a \emph{Penrose inequality}.\footnote{The 
term \emph{Penrose inequality} has also been applied to describe similar bounds for 
the mass at spacelike infinity. See~\cite{mom:ts}.} In particular,
this implies that $M_f$ is strictly positive.
\vskip.5pc
\noindent\emph{Proof.} 
Let $M>M_f$. There exists a point $(u_0,V)\in\mathcal{I}^+$ such 
that $m(u_0,V)\le M$. If $(u',v')\in\mathcal{A}'$, we have
\[
(u',u_0]\times v'\cup u_0\times[v',V)\subset\mathcal{R}.
\]
In particular, integrating the inequalities $\partial_v m\ge0$,
$\partial_u m\le 0$, along these segments, yields
\[
m(u',v')\le M.
\]
But, $m(u',v')=\frac{r}2(u',v')$, since $(u',v')\in\mathcal{A}'$,
thus
\[
r\le 2M.
\]
Since this is true for all $M>M_f$, $r\le 2M_f$.
$\Box$

\subsection{Penrose inequality for the event horizon}
Next, we shall prove the following:
\begin{lemma}
\label{gegovotapenrose}
On $\mathcal{H}$, 
$r\le 2M_f$.
\end{lemma}

\noindent
The above lemma is yet another manifestation of
a \emph{Penrose inequality}. Since $\partial_u r<0$ immediately yields
\[
\sup_{\mathcal{H}} r\ge \sup_{\mathcal{A}'} r,
\]
it follows that
the above lemma, interpreted as a lower bound on $M_f$, is a stronger
statement than the previous.
\vskip.5pc
\noindent\emph{Proof.} Suppose not.  Then there exists a point
$(\tilde{U},\tilde{V})$ on the event horizon such that 
$r(\tilde{U},\tilde{V})=R>2M_f$.
Let $M$ satisfy $\frac{R}2>M>M_f$. It follows that there exists a point
$(u_0,V)\in\mathcal{I}^+$ such that $m(u_0,V)\le M$.
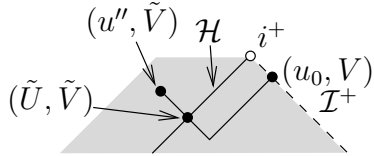
\[ 
\input{snmeia.pstex_t}
\]
Thus,
by the inequalities $\partial_vm\ge0$, $\partial_um\le0$ in 
$J^{-}(\mathcal{I}^+)\cup\mathcal{H}$, it follows that
$m\le M$ in $(J^{-}(\mathcal{I}^+)\cup\mathcal{H})\cap\{u\ge u^0\}$,
in particular, on $\mathcal{H}$. 

It follows now from $(\ref{profavws})$ that 
$r(\tilde{U},v^*)\ge R$, for $v^*\ge \tilde{V}$. In particular,
since $m(\tilde{U},v^*)\le M$, it follows that $1-\frac{2m}r>0$ on
$\mathcal{H}\cap \{v\ge \tilde{V}\}$. 
By Proposition $\ref{toprwto}$, we have in fact
$\mathcal{H}\subset\mathcal{R}$.

By continuity, it now
follows that there exists a $u''$, such that 
$[\tilde{U},u'']\times \tilde{V}\subset\mathcal{R}$,
and $r(u^*,\tilde{V})> R'$, for some $R'<R$ with $1-\frac{2M}{R'}>0$, 
for $u^*\in[\tilde{U},u'']$. 
Consider the set
\begin{eqnarray*}
X&=&[\tilde{U},u'']\times[\tilde{V},V)\cap\mathcal{Q}^+\\
&&\hbox{}\cap 
\{(\bar{u},\bar{v}):r(\tilde{u},\tilde{v})> R'', m(\tilde{u},\tilde{v})<M'
{\rm\ for\ all\ }\tilde{U}\le\tilde{u}\le\bar{u}, 
\tilde{V}\le\tilde{v}\le\bar{v}\}
\end{eqnarray*}
for some $M'>M$, $R''<R'$ such that $1-\frac{2M'}{R''}>0$. By the
continuity of $r$ and $m$,
$X$ is clearly an open subset of 
$[\tilde{U},u'']\times[\tilde{V},V)\cap\mathcal{Q}^+$ 
in the latter set's topology. Moreover, 
since $1-\frac{2m}r>0$ in $X$, it follows that $X\subset\mathcal{R}$. Since
this implies that $\partial_v r\ge 0$, $\partial_u m\le 0$,
it follows that 
$r(\bar{u},\bar{v})>R'$, $m(\bar{u},\bar{v})\le M$, and,
thus $X$ is closed. Since $X$ is clearly connected, it follows
that
\[
X=[\tilde{U},u'']\times[\tilde{V},V)\cap\mathcal{Q}^+.
\]
It follows that $r\ge R'>0$ on 
$[\tilde{U},u'']\times[v,V)\cap\mathcal{Q}^*$,
and thus, by ${\bf \Sigma T}'$,
\[
[\tilde{U},u'']\times[\tilde{V},V)\cap\mathcal{Q}^*=
[\tilde{U},u'']\times[\tilde{V},V)\cap\mathcal{Q}^+.
\]
But the left hand side is the closure of the right hand
side in the topology of $[\tilde{U},u'']\times[\tilde{V},V)$.
Thus, 
\[
[\tilde{U},u'']\times[\tilde{V},V)\cap\mathcal{Q}^+
\]
is an open and closed subset of 
\[
[\tilde{U},u'']\times[\tilde{V},V),
\]
and consequently,
\[
[\tilde{U},u'']\times[\tilde{V},V)\cap\mathcal{Q}^+
=
[\tilde{U},u'']\times[\tilde{V},V),
\]
and moreover,
\[
[u_0,u'']\times[\tilde{V},V)\subset\mathcal{R}.
\]
Integrating $(\ref{m-u})$, noting that both terms on the right
hand side are non-negative in $\mathcal{R}$, 
and that the Hawking mass satisfies
$0\le m\le M$ in $[u_0,u'']\times[\tilde{V},V)$, we obtain the estimate
\begin{equation}
\label{evergeia}
\sup_{\bar{v}\ge \tilde{V}}
\int_{u_0}^{u^*}
{8\pi r^2}T_{uu}\frac{1-\frac{2m}r}{(-\partial_u r)}(\bar{u},\bar{v})
d\bar{u}\le M.
\end{equation}

Consider now the quantity
\[
\frac{\partial_vr}{1-\frac{2m}r}.
\]
This is well defined at $(u_0,v^*)$ for all $v_*\in[\tilde{V},V)$. 
We easily compute the identity
\begin{equation}
\label{k-e3is2}
\partial_u\frac{\partial_vr}{1-\frac{2m}r}
=
\frac{4\pi rT_{uu}}{\partial_ur}\frac{\partial_vr}{1-\frac{2m}r}.
\end{equation}
In view of the bounds
\[
r\ge R',
\]
\[
1-\frac{2m}r\ge1-\frac{2M}{R'}
\]
in $[\tilde{U},u'']\times[\tilde{V},V)$,
$(\ref{evergeia})$ yields
\begin{equation}
\label{movnektimnsn}
\sup_{\bar{v}\ge \tilde{V}}
\int_{u_0}^{u^*}\frac{4\pi rT_{uu}}{(-\partial_ur)}d\bar{u}\le
\frac{M}{2(R'-2M)}
\end{equation}
and thus,
integrating (\ref{k-e3is2}),
\[
\frac{\partial_vr}{1-\frac{2m}r}(u^*,v^*)\ge\exp(-M(2(R'-2M))^{-1})
\frac{\partial_vr}{1-\frac{2m}r}(u_0,v^*).
\]
We obtain immediately
that 
\[
\partial_vr(u^*,v^*)\ge\left(1-\frac{2M}R'\right)^{-1}\exp(-M(2(R'-2M))^{-1})
\partial_vr(u_0,v^*),
\]
and thus, upon integration
$r(u^*,v^*)\to\infty$ as $v^*\to V$, since
$r(u_0,v^*)\to\infty$, i.e.~$(u^*,V)\in\mathcal{I}^+$. 
It follows that $\mathcal{H}$
is not the event horizon after all, a contradiction. $\Box$

\subsection{Christodoulou's completeness condition}
The completeness statement we shall prove is the formulation
of~\cite{chr:givp}. In our context, this takes the following form:
Fix an outgoing null ray $u=u_0$, for $u_0<\tilde{U}$, and
consider the vector field
\[
{\bf X}(u,v)=\frac{\partial_vr(u,v)(1-\frac{2m}r)(u_0,v)\partial_ur(u,v)}
		{\partial_vr(u_0,v)(1-\frac{2m}r)(u,v)
			\partial_ur(u_0,v)}\frac{\bf \partial}{\bf \partial u}
\]
on $J^{-}(\mathcal{I}^+)\cap\mathcal{Q}^+$. 
This vector field is parallel along the outgoing null ray $u=u_0$,
and on all ingoing null rays. We shall show that the affine
length
\[
\int_{u_0}^{\tilde{U}}{\bf X}(u,v)\cdot udu\to\infty 
\]
as $v\to\infty$.

Let $R>2M_f$, and consider the curve $\{r=R\}\cap J^{-}(\mathcal{I}^+)$.
By Lemma \ref{faivomevo}, for sufficiently large $v_0<V$, all ingoing null curves
with $v\ge v_0$ intersect $\{r=R\}\cap J^{-}(\mathcal{I}^+)$ 
at a unique point $(u^*(v),v)$, depending on $v$. 
Let $M$ denote the Bondi mass at $u_0$.
We have
\begin{eqnarray}
\label{teleutaio}
\nonumber
\int_{u_0}^{\tilde{U}}{\bf X}(u,v)\cdot udu
	&\ge&
		\int_{u_0}^{u^*(v)}{\bf X}(u,v)\cdot udu\\
	&\ge&
\nonumber
		\frac{1}{(-\partial_ur)(u_0,v)}
		\int_{u_0}^{u^*(v)}{\exp{\left(\int_{u_0}^u{\frac{4\pi rT_{uu}}
						{\partial_u r}(\bar{u},v)d\bar{u}}
						\right)}
			(-\partial_ur)du}\\
	&\ge&
		\frac{(r(u_0,v)-R)}{(-\partial_ur)(u_0,v)}\exp(-M(2(R-2M))^{-1}),
\end{eqnarray}
where $(\ref{teleutaio})$ follows from
the bound
\[
\int_{u_0}^u{\frac{4\pi rT_{uu}}
						{(-\partial_u r)}(\bar{u},v)d\bar{u}}
\le \frac{M}{2(R-2M)},
\]
which is proven as in $(\ref{movnektimnsn})$.
Since $r(u_0,v)\to\infty$ as $v\to\infty$, to show the theorem,
it suffices to show
that $(-\partial_ur)(u_0,v)$ is uniformly bounded in $v$.

Consider the quantity
\[
\frac{\partial_ur}{1-\frac{2m}r}.
\]
In analogy to $(\ref{k-e3is2})$,
we have
\[
\partial_v
\frac{\partial_ur}{1-\frac{2m}r}=
\frac{4\pi r T_{vv}}{\partial_v r}\frac{\partial_ur}{1-\frac{2m}r},
\]
and thus
\[
\frac{\partial_ur}{1-\frac{2m}r}(u_0,v)=
\exp\left(\int_{v_0}^v{\frac{4\pi r T_{vv}}{\partial_v r}(u_0,\bar{v})d\bar{v}}
\right)\frac{\partial_ur}{1-\frac{2m}r}(u_0,v).
\]
Recalling the defintion of $M$, we can choose $v_0$
such that $1-\frac{r(u_0,v_0)}{2M}>0$. Set $R'=r(u_0,v_0)$.
Then, as in $(\ref{movnektimnsn})$, we have
\[
\int_{v_0}^{v}\frac{4\pi rT_{vv}}{(\partial_vr)}d\bar{v}\le
\frac{M}{2(R'-2M)},
\]
and thus,
\[
-\partial_ur(u_0,v)\le \left(1-\frac{R'}{2M}\right)^{-1}
\exp\left(M(2(R'-2M))^{-1}\right)
\]
for $v\ge v_0$. This completes the proof. $\Box$

\section{Remarks}
It is clear that in the above theorem, 
we only used the condition $\mathcal{A}\cup\mathcal{T}\ne\emptyset$ to
infer $\mathcal{Q}\setminus\overline{J^{-}(\mathcal{I}^+)}
\ne\emptyset$. Thus, it follows that we
have in fact proven
\begin{theorem}
\label{allo}
If 
$\mathcal{Q}\setminus\overline{J^-(\mathcal{I^+})}\ne\emptyset$, 
then $\mathcal{I}^+$
is complete and Lemma~\ref{gegovotapenrose} holds.
\end{theorem}

Another point is worth mentioning. 
It turns out that all statements of this paper except the
positivity of mass in $\mathcal{R}$ of Proposition~\ref{toprwto} hold equally well
if ${\bf A}'$ is modified to

\begin{enumerate}
\item[${\bf \tilde A}'$]
In ${\bf A}'$, let
$\Gamma$ now be assumed to be an ingoing null segment,
let the condition $r=0$ on $\Gamma$ be dropped, while let the
condition 
\begin{equation}
\label{veasuv9nkn}
1-\frac{2m}r>0
\end{equation}
on $S$ be added. 
\end{enumerate}
(Positivity of mass in $\mathcal{R}$ holds if it is assumed
on $\Gamma$.)

Given a spherically symmetric initial data for a ``reasonable''
Einstein-matter system, but where the data now has
$2$ asymptotically flat ends, let $S$ be a connected
piece of the quotient of one of the ends, so that
$(\ref{veasuv9nkn})$ holds, and such that the inward expansion is
everywhere negative along $S$, let $p$ be the endpoint of
$S$, and let $\Gamma$ be the ingoing null curve in the quotient
future development, emanating from $p$. It is clear that
${\bf \tilde A}'$ holds for $\mathcal{Q}^+=J^+(\Gamma\cup S)$.

In particular, ${\bf \tilde A}'$, ${\bf B}'$--${\bf \Sigma T}'$ hold
for the region $\mathcal{Q}^+$ defined as above for the
development of the Einstein-Maxwell-scalar field system
studied in~\cite{md:si,md:cbh,mi:mazi}.

Finally, it might be useful 
to point out what we have \emph{not} shown.
We have not shown that $i^+\in\overline{\mathcal{A}'}$, and we have
not shown that $\sup_{\mathcal{H}}r=2\sup_{\mathcal{H}}m$. 
Both these statement
are true, however, in the case of a self-gravitating scalar field.

\end{document}